\documentclass[]{spie}  %>>> use for US letter paper
\usepackage{amsmath,amsfonts,amssymb}
\usepackage{graphicx}
\usepackage[colorlinks=true, allcolors=blue]{hyperref}

\def\arcdeg{\mbox{$^\circ$}}%
\def\arcsec{\mbox{$^{\prime\prime}$}}%

\title{Ground calibration plans for the AXIS high speed camera}

\author[a]{Catherine E.\ Grant}
\author[a]{Eric D.\ Miller}
\author[a]{Marshall W.\ Bautz}
\author[a]{Jill Juneau}
\author[a]{Beverly J.\ LaMarr}
\author[a]{Andrew Malonis}
\author[a]{Gregory Y.\ Prigozhin}
\author[b]{Christopher W.\ Leitz}
\author[c,d,e]{Steven W.\ Allen}
\author[c]{Tanmoy Chattopadhyay}
\author[c]{Sven Herrmann}
\author[c]{R.\ Glenn Morris}
\author[c,d]{Abigail Y.\ Pan}
\author[c]{Artem Poliszczuk}
\author[c,d]{Haley R.\ Stueber}
\author[c]{Daniel R.\ Wilkins}
\affil[a]{Kavli Institute for Astrophysics and Space Research, Massachusetts Institute of Technology, Cambridge, MA, USA}
\affil[b]{Lincoln Laboratory, Massachusetts Institute of Technology, Lexington, MA, USA}
\affil[c]{Kavli Institute of Astrophysics and Cosmology, Stanford University, Stanford, CA, USA}
\affil[d]{Department of Physics, Stanford University, Stanford, CA, USA}
\affil[e]{SLAC National Accelerator Laboratory, Menlo Park, CA USA}

\authorinfo{Further author information: C.\ E.\ Grant, E-mail: cgrant@mit.edu}

\begin{document} 
\maketitle

\begin{abstract}
The Advanced X-ray Imaging Satellite (AXIS), an astrophysics NASA probe mission currently in phase A, will provide high-throughput, high-spatial resolution X-ray imaging in the 0.3 to 10~keV band. We report on the notional ground calibration plan for the High Speed Camera on AXIS, which is being developed at the MIT Kavli Institute for Astrophysics and Space Research using state-of-the-art CCDs provided by MIT Lincoln Laboratory in combination with an integrated, high-speed ASIC readout chip from Stanford University. AXIS camera ground calibration draws on previous experience with X-ray CCD focal plans, in particular Chandra/ACIS and Suzaku/XIS, utilizing mono-energetic X-ray line sources to measure spectral resolution and quantum efficiency. Relative quantum efficiency of the CCDs will be measured against an sCMOS device, with known absolute calibration from synchrotron measurements. We walk through the envisioned CCD calibration pipeline and we discuss the observatory-level science and calibration requirements and how they inform the camera calibration.
\end{abstract}


\section{INTRODUCTION}
\label{sec:intro}

The Advanced X-ray Imaging Satellite (AXIS) is an astrophysics NASA probe-class mission currently in phase A with a planned launch of 2032.\cite{2024SPIE_AXIS}  AXIS provides high-throughput, high-spatial-resolution X-ray imaging spectroscopy, with an agile spacecraft for rapid transient response, addressing many of the key science priorities identified by the Astro2020 decadal survey.\cite{Astro2020}  AXIS is a single instrument observatory, combining arc-second quality, large effective area, lightweight silicon mirrors with a state-of-the-art X-ray CCD camera.

The AXIS High Speed Camera is being developed at the MIT Kavli Institute for Astrophysics and Space Research (MKI) in collaboration with Stanford University's Kavli Institute for Particle Astrophysics and Cosmology (KIPAC), using CCDs provided by MIT Lincoln Laboratory (MIT/LL)\cite{Leitz2025_SPIE}.  The high throughput of the mirror requires fast CCD readout, enabled by a high-speed low-power application-specific integrated circuit (ASIC), the Multi-Channel Readout Chip (MCRC) developed by Stanford.\cite{2020SPIE_MCRC,2024SPIE_MCRC} A more detailed description of the AXIS camera, and the development and testing of the CCDs and ASICs, can be found in Ref.~\citenum{2023SPIE_AXIScamera} and \citenum{2025SPIE_AXIScamera}.

Science results from any instrument require accurate knowledge of the performance characteristics of the instrument itself. A fraction of on-orbit observation time is generally dedicated to ``calibration", measuring and monitoring changes of all aspects of the observatory (optics, filters, detectors, etc.).  Observing time for calibration is limited, with priority given to science targets, and calibration targets are often multi-component time-varying astrophysical sources which complicates analysis.\cite{2015JATIS_IACHEC}  Ground calibration measurements can provide a strong baseline understanding of instrument performance before launch, with higher statistics, well understood X-ray sources, and a wider range of operating parameters. Quantum efficiency measurements are particularly difficult and important given the need for stable calibration sources and the complex response function of X-ray detectors. Ground calibration can better inform physics-based models of system performance, which can then be adjusted as the system ages and changes on orbit. Scheduling sufficient time and resources to ground calibration is crucial for efficient and successful science analysis after launch.   

During the development phase, MIT/LL is producing AXIS prototype CCDs, both front- and back-illuminated, pairing them with Stanford-designed ASICs, and delivering them to our dual test facilities at MKI's X-ray Detector Lab\footnote{\url{https://sites.mit.edu/xraydetectorlab/}} and the X-ray Astronomy and Observational Cosmology (XOC) group\footnote{\url{https://xoc.stanford.edu}} at Stanford. Initially these will be in test packages, as described in Ref.~\citenum{2025SPIE_AXIScamera} and \citenum{2025SPIE_Pan}, and later in flight-like packages. MKI and Stanford groups will test the use of multiple readouts, measure noise, tune voltages, and learn as much about the packaged devices as possible to identify any issues in design, test hardware, or procedures so they can be addressed before the flight CCD screening and calibration process begins.

As the AXIS flight candidate CCDs are manufactured by MIT/LL and delivered to MKI, they will go through a screening and calibration process before the flight devices are assembled into the camera focal plane. The full camera team, from MKI, Stanford, and MIT/LL, will support this effort. Once the camera is assembled, a radioactive $^{55}$Fe source in the camera door will be used for aliveness and performance verification, but no further large-scale ground calibration activities are planned. (Calibration of the X-ray mirrors will be done separately by the team at GSFC and is beyond the scope of this paper.) After presenting the basic characteristics of the AXIS camera, this paper will outline the notional ground calibration process and the products that will be generated. ACIS on the Chandra X-ray Observatory and XIS on Suzaku also went through CCD-level calibration at MKI\cite{1998SPIE_ACISGCal,2005NIMPA_XISGCal}; the AXIS ground calibration plan follows a similar path to both previous successful efforts. The figures in this work are examples of similar analysis done on an earlier generation of MIT/LL CCD incorporating AXIS technology, but not the CCD planned for AXIS.

\section{AXIS Camera}
\label{sec:ccd}

\begin{figure}[t]
\begin{center}
\includegraphics[width=0.65\linewidth]{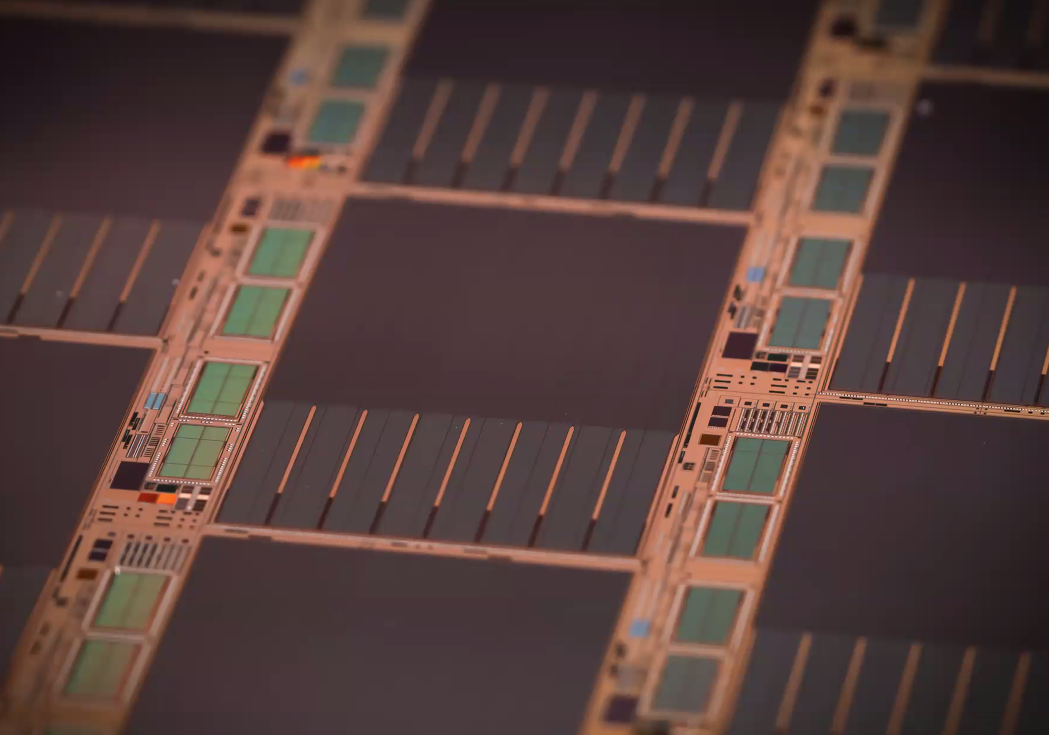}
\includegraphics[width=0.3\linewidth]{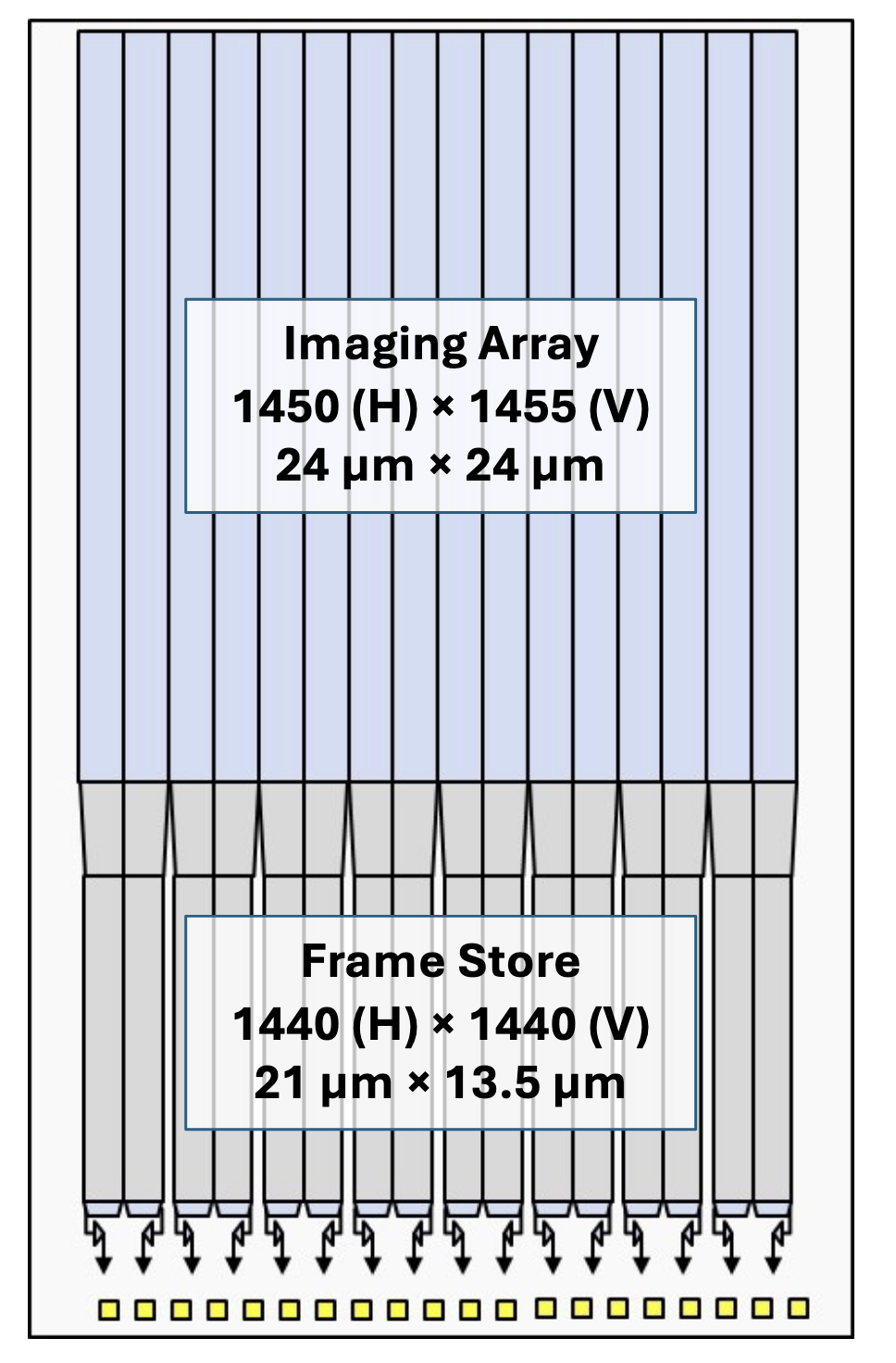}
\vspace{0.1in}
\caption{(left) A photograph of an AXIS CCD wafer fabricated by MIT/LL.  Each wafer contains seven complete CCID100 CCDs.  (right) A diagram of a CCID100 device.  The 1440 by 1440 pixel active imaging array is readout through 16 output ports.}
\label{fig:ccid100}
\end{center}
\end{figure}

The heart of the AXIS High Speed Camera is a set of four X-ray CCDs fabricated by MIT Lincoln Laboratory.  These state-of-the-art CCDs, denoted CCID100\cite{Leitz2025_SPIE}, build on more than three decades of successful X-ray CCD development and deployment in space observatories (ASCA, HETE-2, Chandra, Suzaku).\cite{2024SPIE_Donlon} Characteristics of the AXIS sensor are shown in Figure~\ref{fig:ccid100} and Table~\ref{tab:ccid100}.  They are back-illuminated and 100~$\mu$m thick, fully depleted, with good response in the full energy range of 0.3 to 10~keV. To meet mission requirements and avoid photon pileup, these are large devices with much faster pixel rates than previous missions. The fast readout is achieved both by advances in the readout sense nodes, with readout rates $>$~20 times those in heritage devices, and increasing the number of readout nodes per device. This also necessitates the use of the low-power, small-footprint MCRC ASIC for off-chip signal amplification.  During ground calibration, the fast readout time allows higher incident photon rates without significant pileup, however the AXIS sensors are read out through 16 output ports, and each one will require separate gain and response calibration.  

\begin{table}
\caption{Characteristics of the AXIS CCID100 devices}    
\label{tab:ccid100}
\begin{center}
    \begin{tabular}{|l|l|}
\hline\hline
%Characteristic &Value \\ \hline
Detectors &Back-illuminated frame transfer CCDs\\ \hline
Format &1440 $\times$ 1440 pixels active area \\ \hline
Image area pixel size &24 $\mu$m (0.55\arcsec) \\\hline
Output ports & 16\: 2-stage pJFETs \\\hline
Radiation tolerance features &Trough, charge injection \\\hline
Detector thickness &100 $\mu$m, operated fully depleted \\\hline
Back surface  &Molecular Beam Epitaxy passivation + \\
&on-chip Al optical blocking filter\\\hline
System read noise &$\leq$ 3 $e^-$ RMS requirement \\\hline
Full frame readout time & $\leq$ 200 msec requirement\\ \hline
Focal plane temperature &$-110 \pm 0.1$ \arcdeg C \\\hline
\hline
\end{tabular}
\end{center}
\end{table}

The AXIS CCDs are capable of charge injection, which can be used to measure and mitigate charge transfer inefficiency (CTI).\cite{2009PASJ_SuzakuCI}  Flight devices are expected to initially have very low CTI\cite{LaMarretal2022b,2024SPIE_LaMarr} and the use of charge injection is not planned for normal flight operation, as it entails a small loss of effective area and an increase in noise.  Eventually, after many years exposed to the space radiation environment, charge injection may become an option for better spectral performance.  The ground screening and calibration process will include some basic checks of the charge injection capability of the CCDs which will provide experience for any future use.

As CCDs are also excellent detectors in the UV/optical bands, blocking filters are required.  The AXIS camera is planned to have two: a freestanding filter, consisting of aluminum and polyimide on a stainless steel mesh, and an on-chip filter of aluminum directly deposited on the CCD entrance window. For CCD ground calibration, the on-chip filter will be in place, and the additional absorption will be included in the CCD quantum efficiency measurement.  The freestanding filter will be calibrated independently from the CCDs at a synchrotron, before it is installed in the camera.

The AXIS flight camera includes two Front-End Electronics (FEE) boxes which control and read out each CCD+ASIC pair, digitizing the pixel stream. The flight FEE boards will not be available for ground calibration at the CCD-level in the lab, but after integration of the full camera system, noise and spectral performance will be measured and monitored using the $^{55}$Fe door source.  Prototype FEE electronics boards will be incorporated into the ground calibration process if and when they are available.

The digitized pixel stream from the FEE is then routed to the Back-End Electronics (BEE) box where the full frame data is processed to select candidate X-ray events and reduce the required telemetry bandwidth. For ground calibration there is sufficient storage space to save all the original full frame images. Event finding and filtering is done using standard computer scripts, including the capability to reprocess the data with different settings as needed.

\section{SCREENING}
\label{sec:screen}

Initial CCD qualification begins as devices are fabricated at MIT/LL. Extensive in-process monitoring will demonstrate functional performance at room temperature. Candidate acceptable devices will be integrated with Stanford-provided ASICs and delivered in a flight package to MKI's X-ray Detector Lab for further screening and performance calibration.

The packaged devices will be installed in a dedicated screening vacuum chamber, with a cryostat capable of maintaining stable CCD temperatures down to $-120 \arcdeg$~C. An Archon\footnote{\url{http://www.sta-inc.net/archon}} controller will provide CCD bias and clock voltages and digitize the pixel stream, in the same manner as the FEE will in the flight instrument. As performance can be dependent on temperature, CCD frametime, and clocking speed, the nominal flight values will be used, except where noted. The chamber will be equipped with an insertable $^{55}$Fe source, providing X-ray lines at Mn-K$\alpha$ and K$\beta$ (5.9 and 6.4~keV). In the interests of time, if at any point the device is clearly not a good candidate for flight, screening will stop and proceed to the next candidate. 

\begin{figure}[t]
\begin{center}
\includegraphics[width=.45\linewidth]{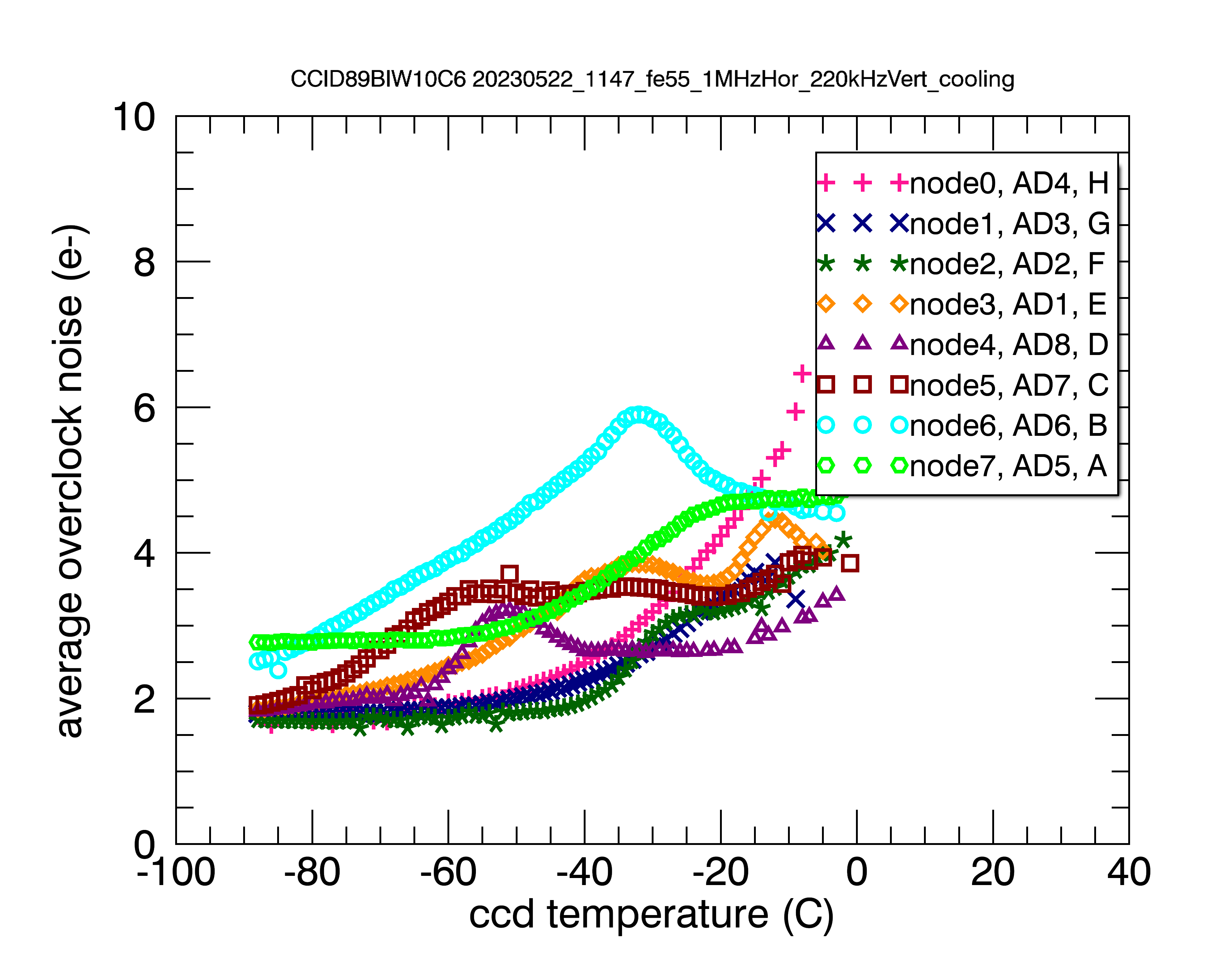}
%{20230526_2MHzHor_220kHzVert_p_noise_temp_e.png}
\includegraphics[width=0.49\linewidth]{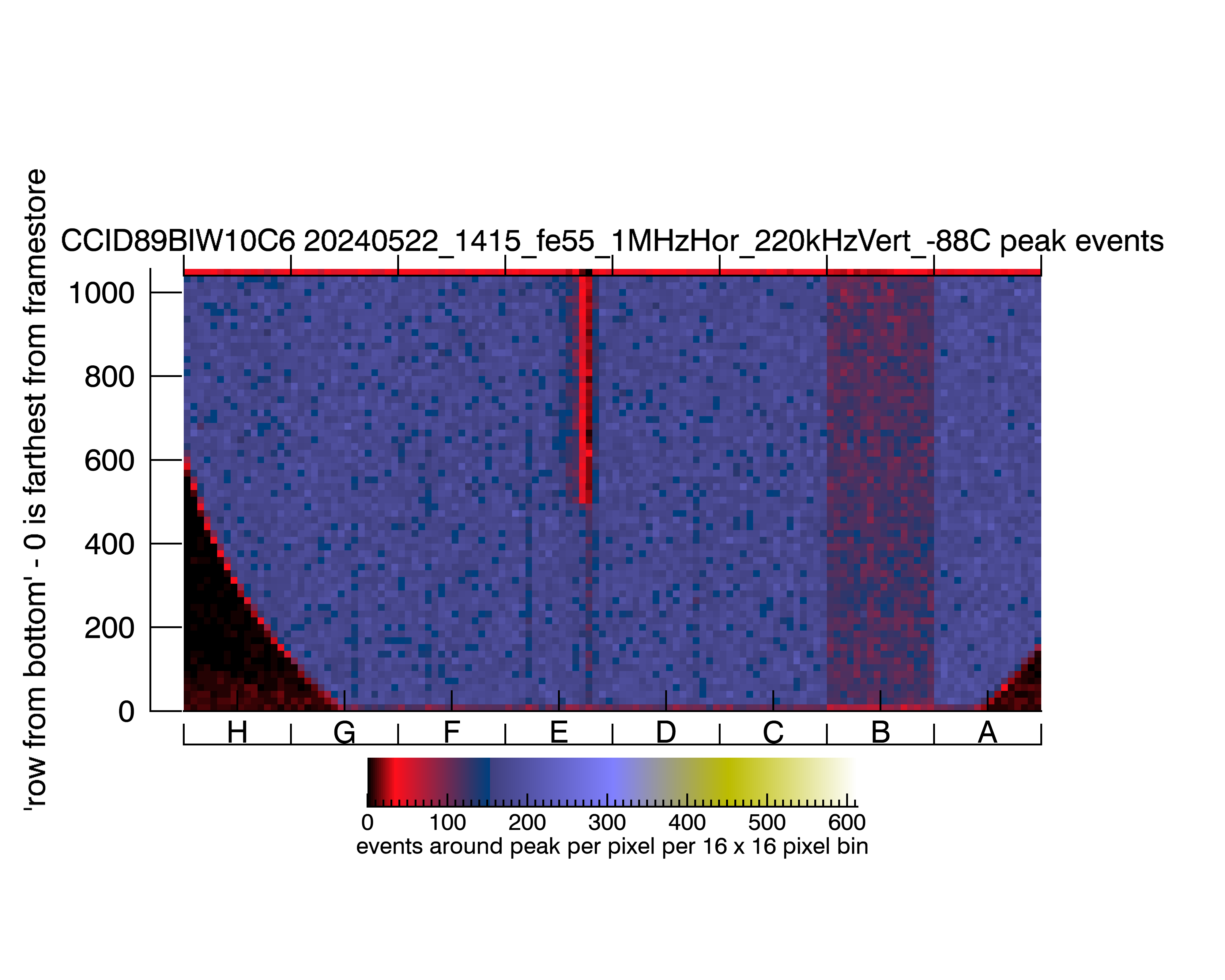}
\includegraphics[width=.45\linewidth]{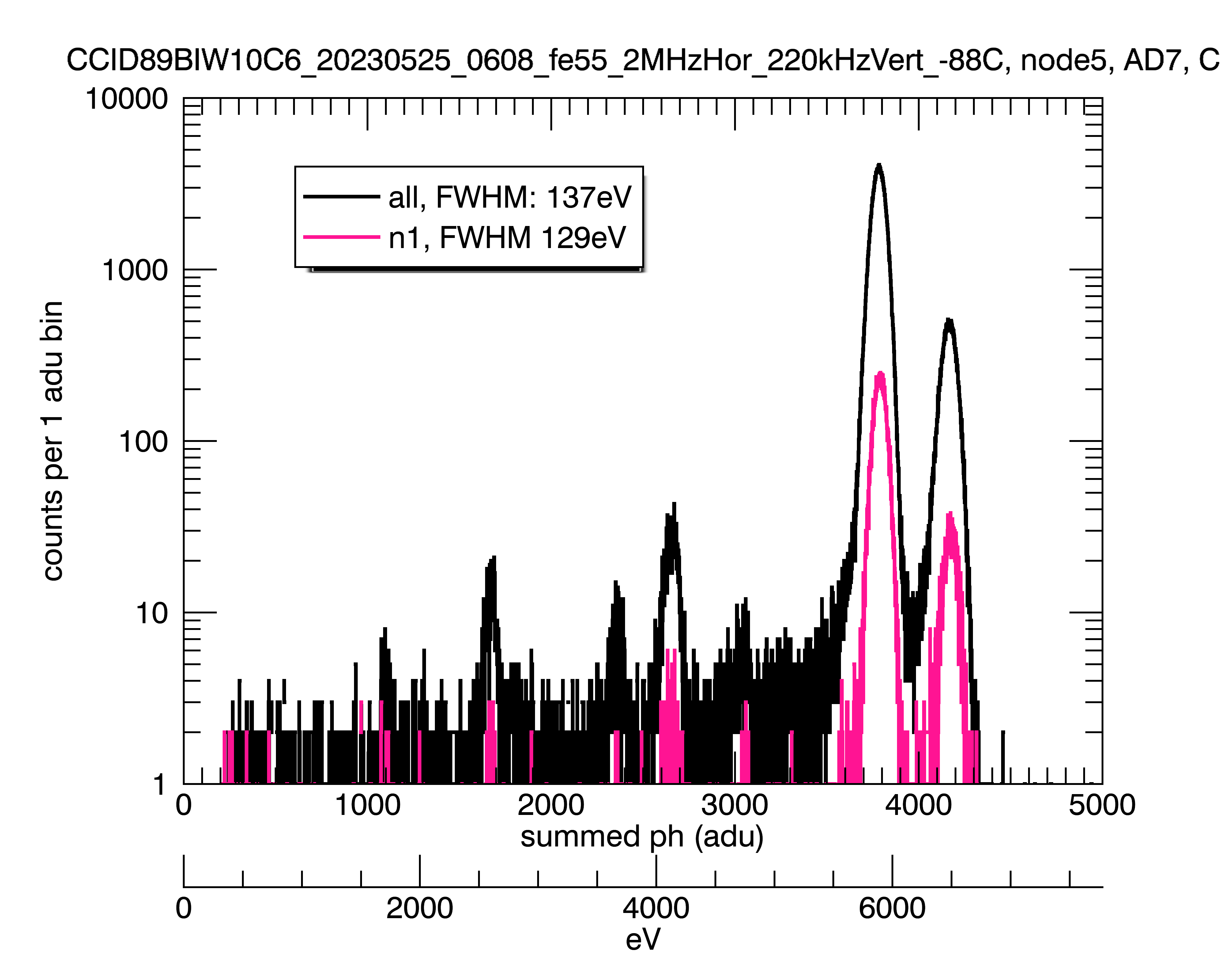}
\includegraphics[width=0.49\linewidth]{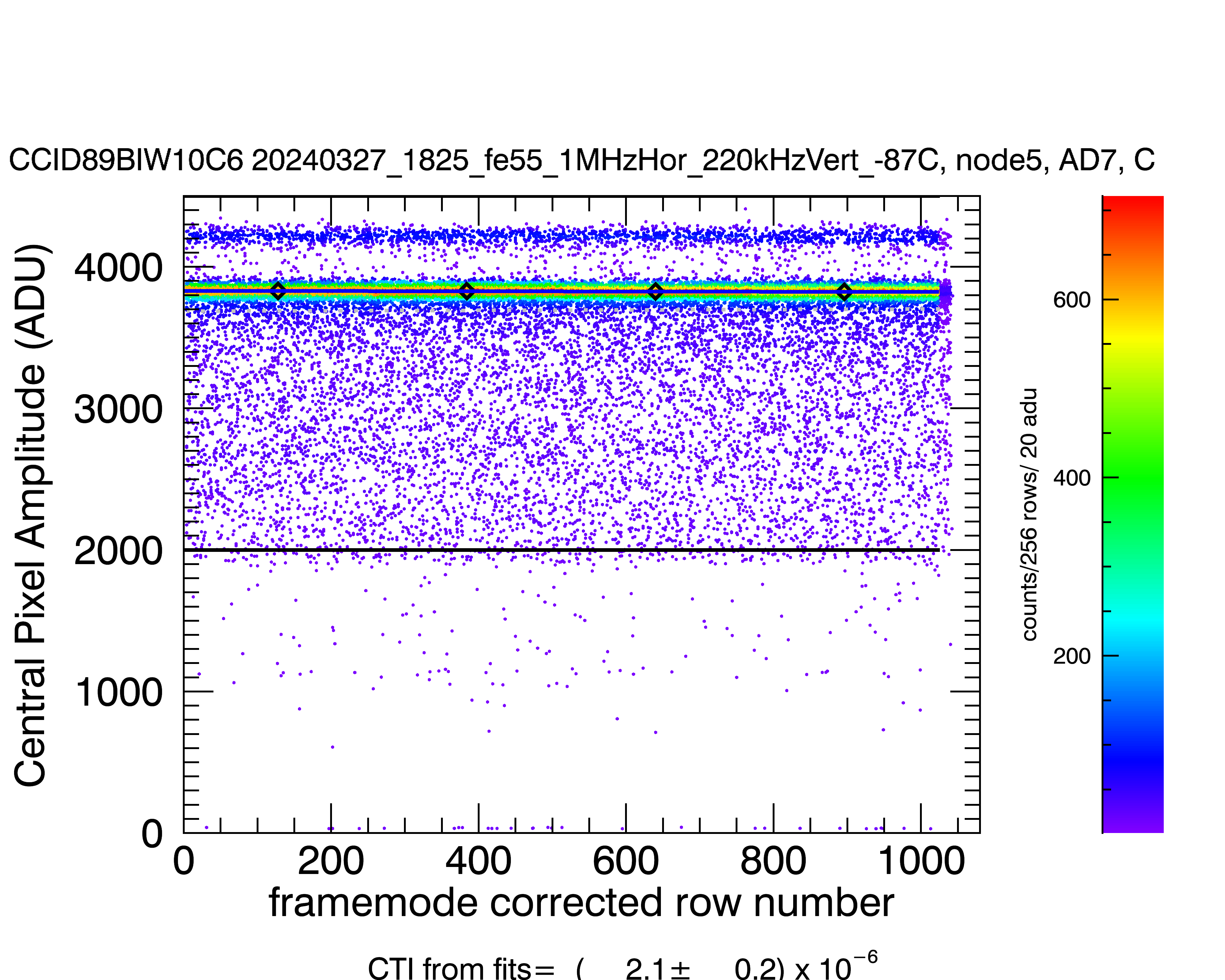}
\end{center}
\caption{An example of the type of data and analysis that will be used to screen AXIS CCDs.  These data are from an earlier generation of MIT/LL CCD incorporating AXIS technology, not the CCID100.
\textbf{Upper left:} Readout noise in $e^-$ RMS as a function of temperature for all output nodes. All of the eight nodes have lower noise than the required $<$~3~$e^-$~RMS at $-90\arcdeg$~C, the lowest temperature measured in this particular test. 
\textbf{Upper right:} An image of the $^{55}$Fe events on the CCD.  The curvature seen in the corners is due to a misalignment of the CCD with the source.  The red stripe near the center is a cosmetic defect, a group of partial hot columns. 
\textbf{Lower left:} $^{55}$Fe spectrum from a representative node of the same CCD for single pixel events (pink) and all events (black). The bright lines of Mn K$\alpha$ (5.9~keV) and K$\beta$ (6.4~keV) are easily resolved and meet the AXIS requirements for spectral resolution at 6 keV ($\leq$150~eV FWHM). 
\textbf{Lower right:} Parallel CTI for the same representative node of the same CCD. Each point in each plot represents a single X-ray event.  The color coding indicates the number of events in the row and amplitude bin.  The diamonds are the center of a Gaussian fit to the center pixel amplitudes.  The dashed line is a linear fit to the center of the gaussian fits.
(From Ref.~\citenum{2023SPIE_AXIScamera} and \citenum{2024SPIE_LaMarr})}
\label{fig:89noisefe55}
\end{figure} 

After initial cooling to the nominal operating temperature of $-110\arcdeg$ C, a brief tuning process will determine the optimal values for clocking voltages, as it is expected that readout noise performance may vary slightly between devices and readout nodes. An example tuning procedure is discussed in Ref.~\citenum{2025SPIE_Stueber}. These settings can then be transferred to operation with the FEE after assembly of the camera.

Screening begins with the $^{55}$Fe source, which fully illuminates all 16 output nodes of the CCD. The readout noise is computed during a temperature scan while the CCD is cooled down (or warmed up) by finding the RMS variation of serial `overlock' pixels, converted to $e^-$ by using the gain measured at Mn-K$\alpha$.  At the nominal operating temperature, $-110\arcdeg$~C, modest exposures are sufficient to reveal cosmetic defects, and measure gain, FWHM, and charge transfer inefficiency.  Examples of the type of data used for screening is shown in Figure~\ref{fig:89noisefe55}.  Readout noise should be below the requirement of 3~$e^-$~RMS, cosmetic defects should be minimal, the spectral line profiles should be sharp and well separated, with Gaussian FWHM below the AXIS spectral resolution requirement of 150~eV, and CTI should be close to zero, with typical values of a few~$\times 10^{-6}$ or lower.

\begin{figure}[t]
\begin{center}
\includegraphics[width=5in]{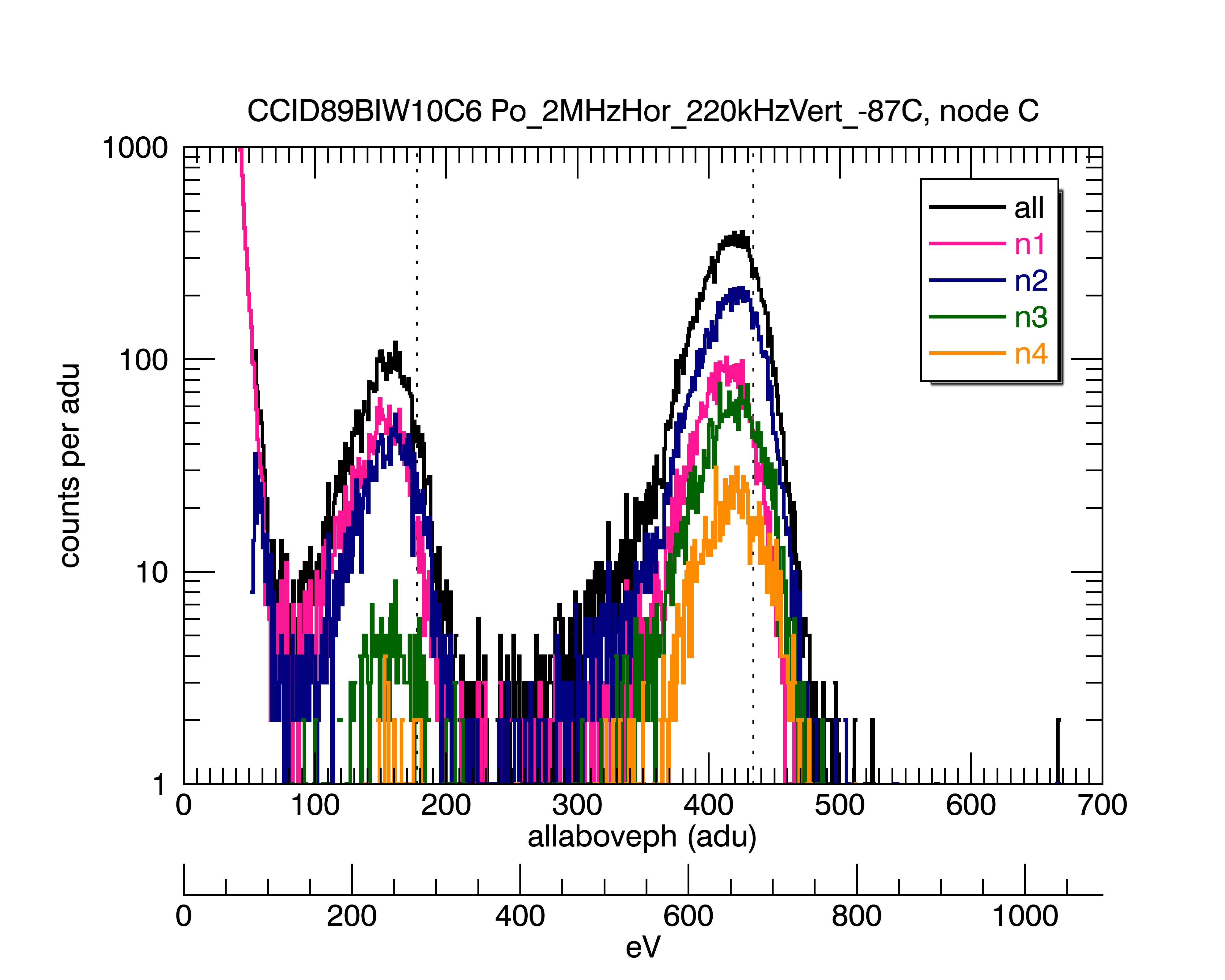}
\end{center}
\caption{An example spectrum of the C-K (277~eV) and F-K (677~eV) fluorescence lines produced by a $^{210}$Po source with a Teflon target. The fluorescence lines are well separated and suitable for calibration use. Spectra for events of different pixel multiplicities are shown in different colors (see legend). (From Ref.~\citenum{2023SPIE_AXIScamera}.)}
\label{fig:89po}
\end{figure} 

Low energy response can then be explored using a radioactive $^{210}$Po source with a Teflon target producing lines of C-K (277~eV) and F-K (677~eV). Figure~\ref{fig:89po} shows an example spectrum from the Polonium source in our lab. Except for the well-measured radioactive decay, the Polonium source has the advantage over a powered source of being very stable with time. Low energy performance is much more sensitive to problems with the back surface of the CCD and issues with depletion than that measured at $^{55}$Fe.  Poor charge collection would show up as a non-Gaussian low energy feature in the spectral line.  Event pixel multiplicity can also indicate problems with charge collection or transfer.  A small low-energy hump is apparent in the figure, but it is orders of magnitude below the peak and would not be considered problematic.

Finally, the candidate devices will be assessed for dark current.  At the nominal focal plane temperature of $-110\arcdeg$~C, the dark current should be negligible and difficult/impossible to measure. We will confirm an upper limit on dark current at $-110\arcdeg$~C and then do a second measurement at $-60\arcdeg$~C where dark current may be more visible.  In both cases the measurement requires short data sets at two frametimes.  As the nominal frametime of 200~msec is quite short, a better measurement may require both frametimes to be longer than the nominal value.

The performance of screened devices is then compared to AXIS requirements for readout noise ($<$~3~$e^-$~RMS) and FWHM at 6~keV and 0.5~keV ($<150$ and $<70$~eV). Cosmetic defects, CTI, and dark current should be minimal.  Non-Gaussian low-energy features should also be small. Devices that meet AXIS performance requirements and are suitable candidates for flight proceed on to the next step and will be more fully characterized.

\section{CALIBRATION}
\label{sec:cal}

\begin{figure}
    \begin{center}
    \includegraphics[width=5in]{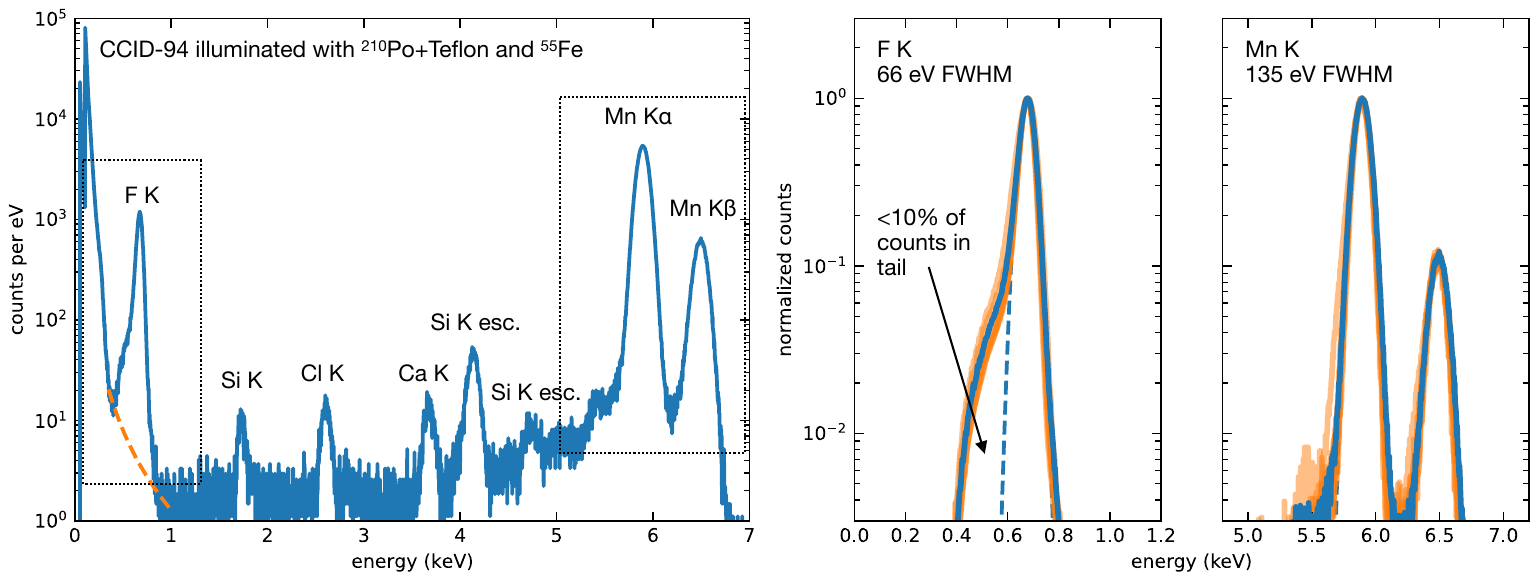}
    \end{center}
\caption{An example spectrum demonstrating the off peak structures in real CCD data.  In this case, the X-ray illumination is lines of C-K (277~eV), F-K (677~eV), Mn-K$\alpha$ (5.9~keV) and Mn-K$\beta$ (6.4~keV).  The Cl and Ca features are from materials in the chamber, but the Si-K fluorescence, Si-K escape lines, and the non-Gaussian features below the main peaks all need to be incorporated in the spectral response file. (From Ref.~\citenum{2023SPIE_AXIScamera})}
\label{fig:94spec}
\end{figure}

Scientific analysis of X-ray CCD event lists requires knowledge of the spectral response and the effective area, both as a function of energy and sky and detector position.  These are generally incorporated into observation specific calibration files, the Redistribution Matrix File (RMF) and the Ancillary Response File (ARF), which are used in astrophysical model fitting.  The RMF maps how mono-energetic X-rays are transformed by the detector into the digitized pulseheight profile.  Gaussian fitting and FWHM is often used to characterize detector performance but the RMF also requires knowledge of the full distribution, including fluorescence and escape peaks, and low-energy structures.  Figure~\ref{fig:94spec} is an example spectrum showing off peak structures in real CCD data. The ARF then contains the effective area as a function of energy at the position of the astrophysical source of interest.  It includes mirror effective area and vignetting, transmission of the freestanding blocking filter, and the quantum efficiency of the detector system. Quantum efficiency is the energy-dependent transmission through any absorbing layers (the on-chip Al optical blocking filter and the CCD surface layers) combined with the absorption probability in the depleted Silicon. 

It is not feasible to use a mono-energetic X-ray source to measure spectral response and quantum efficiency on a fine enough energy scale to create ARFs and RMFs from calibration data alone.  Instead it is necessary to create a physics-based model for the detector system, collect data at a smaller number of energies, then adjust model parameters to best fit the detector performance.  In some cases, calibration files can be populated using careful interpolation between the measurements.

For quantum efficiency, the simplest model is a slab and stop model, using standard absorption/transmission curves for each material including any edge structures. The layer thicknesses are then free parameters fit to the calibration data.  The spectral response requires a more complicated model, including the drift and diffusion of the electron cloud, and surface and pixelization losses.  Such a model is being developed by the MKI detector group; this model compared to laboratory data is shown in Ref.~\citenum{2022JATIS_diffusion}, \citenum{2022SPIE_diffusion} and \citenum{2023SPIE_AXIScamera}.

Precise measurement of the spectral response and quantum efficiency requires much longer exposures than used in device screening. The rule of thumb used in ground calibration of the ACIS CCDs was to accumulate at least 10,000 events in each X-ray line and each region of interest.  For ACIS, the regions were 32 pixel square, which matched the expected size of the spacecraft dither. The size of the AXIS dither box has not yet been formalized, but it is likely to be larger to better cover the gaps between CCDs. For well behaved devices with minimal CTI, the spectral response and quantum efficiency should be reasonably uniform, except for small gain shifts between readout amplifiers, so larger regions can be utilized in calibration.  The source count rate should ideally be adjusted to be as high as possible without causing significant event pileup. Both these factors should increase the efficiency of this calibration pipeline in comparison to ACIS.

The calibration setup is not dissimilar to that used for screening.  The flight candidate packaged device will be installed in a dedicated calibration vacuum chamber, with a cryostat maintaining a CCD temperature of $-110\arcdeg$~C.  If prototype FEE boards are available at time calibration starts, they will be used to control and readout the CCD+ASIC, otherwise, an Archon controller will again be used.  Nominal flight values will be used for CCD clocking and frametime.

The X-ray line sources need to cover the bandpass of interest for AXIS, 0.3 to 10~keV. A notional set of energies for AXIS calibration is C-K (277~eV), O-K (525~eV), Al-K (1.5~keV), Mn-K (5.9~keV), and Cu-K (8.0~keV).  The lowest energy lines (C-K and O-K) can be produced by an In-Focus Monochromator (IFM)\cite{1990ApOpt_IFM} that uses grazing incidence reflection gratings to produce clean monochromatic lines from 175~eV to 1.5~keV. The higher energy lines can be produced by standard fluorescence targets and a commercial X-ray source.  An insertable $^{55}$Fe source will also be available.

\begin{figure}
    \begin{center}
    \includegraphics[width=4.8in]{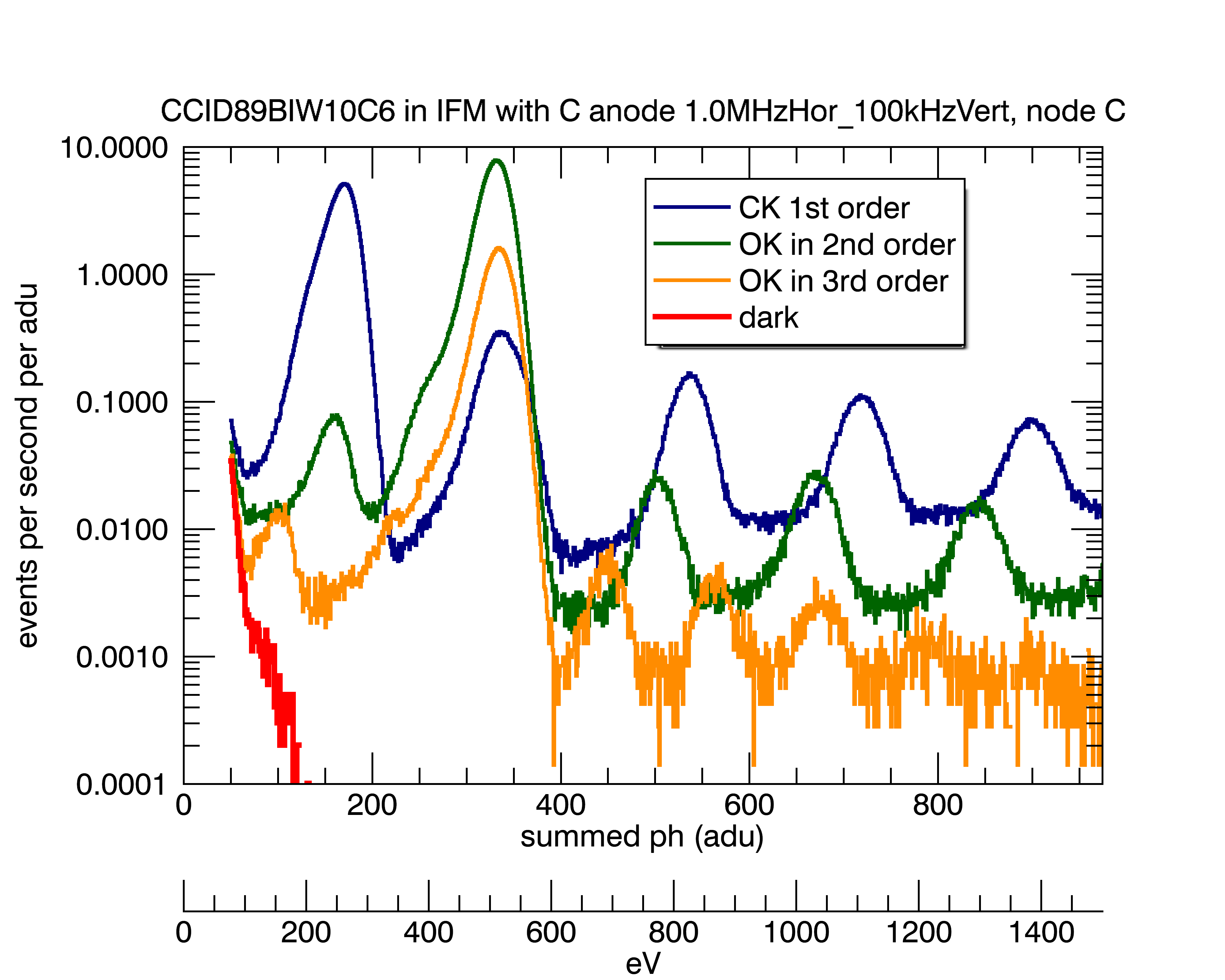}
    \includegraphics[width=0.4\linewidth]{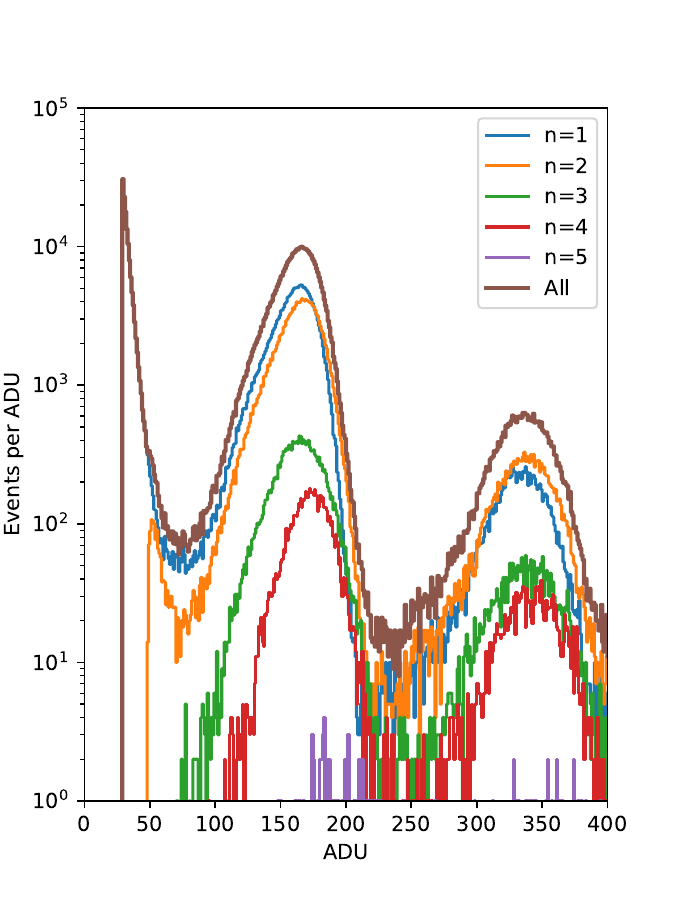}
    \includegraphics[width=0.4\linewidth]{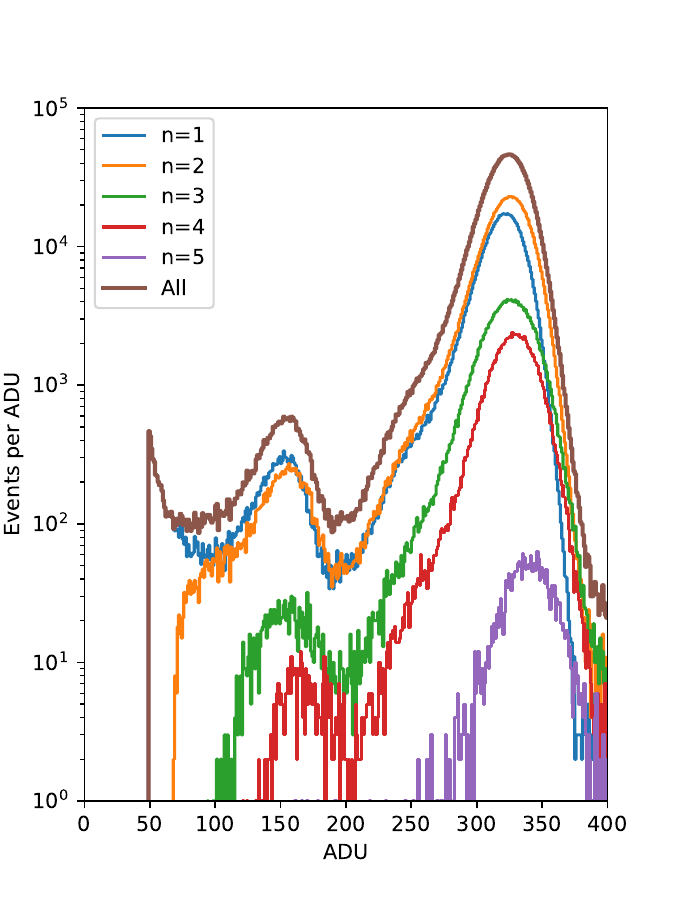}
    \end{center}
\caption{Low energy spectra from the IFM. \textbf{Upper:} Each color shows the spectrum from a different setting of the IFM.  The most prominent peaks are lines from the anode.  The smaller peaks are continuum from the anode passing through other orders of the gratings.  A range of energies from 175~eV to 1.5~keV are possible with this setup. \textbf{Lower left:} O-K (525~eV) in 2nd order. \textbf{Lower right:} C-K (277~eV) in 1st order. Spectra for events of different pixel multiplicities are shown in different colors.  (From Ref.~\citenum{2024SPIE_LaMarr} and \citenum{2024SPIE_Bautz})}
\label{fig:spec}
\end{figure}

Figure~\ref{fig:spec} (upper) shows a composite of spectra taken with different settings of the IFM. The IFM settings can be tuned to put one of the main fluorescence lines of the anode in the first, second, or third order of the grating.  Additional peaks are from continuum going through the grating in other orders, which can also be used for performance measurement, although at lower statistical quality. A closer look at the region around C-K and O-K as seen by the IFM is shown in Figure~\ref{fig:spec} (lower). 

\subsection{Quantum Efficiency Reference Detector}
\label{sec:qe}

Arguably the most difficult parameter to calibrate in any instrument is the effective area, as is requires not just high quality measurements of a source by an observatory, but also knowledge of the absolute flux expected from the source.  On the ground the problem is not that much easier as most sources of X-ray flux in the lab are not absolutely calibrated.  The solution is to use a reference device, which is well calibrated using a source that is completely understood.  Then the test devices can be compared to the reference device and the relative quantum efficiency can be transformed into an absolute quantum efficiency.

Past calibration efforts for Chandra/ACIS and Suzaku/XIS used a MIT/LL CCD similar to the test devices as a reference detector\cite{1997ITNS_ACISBESSY, 2004SPIE_XIScal}.  The reference detectors were absolutely calibrated at the PTB beamline at BESSY in Berlin, using undispersed synchrotron radiation.  The calibration chambers contained a translation stage, which would periodically put the test or the reference CCD in the X-ray source beam.  This system worked well, but has some drawbacks.  Most notably is that the chamber needs to be large, as it needs to fit two CCDs, the translation stage and cooling mechanisms.  For AXIS, where the CCD+ASIC is even larger than ACIS and XIS, having two detectors in the chamber simultaneously is prohibitive.

We have decided to instead use an sCMOS device as a reference detector, and have plans to take such a device to the BESSY-II synchrotron to prepare it for service. We have been investigating the Sony STARVIS family of CMOS devices and are characterizing them in the soft X-ray.\cite{2024JATIS_BenIMX,2024SPIE_Heine}  While they were not designed with X-ray detection in mind, the spectroscopic quality is excellent, with symmetric Gaussian line profiles and near Fano resolution. That simplifies one aspect of transferring absolute calibration, since it's quite simple to measure the flux in a line. Like many sCMOS devices, they are thin and have a microlens layer on the surface which limit the quantum efficiency at high and low energies. Our ongoing program is investigating the best methods to remove the microlenses which improves the quantum efficiency at low energies.

The Sony devices are small and require more modest cooling than a CCD, operating at $\sim$20\arcdeg~C, making accommodation in the chamber less complicated. We are designing a beamline addition putting the sCMOS on a translation stage that can move in front of the CCD to intersect the X-ray beam, then retract completely to allow the test device to see the beam.  The test and reference devices will be at different distances to the source, which will need to be carefully measured, and we have plans for methods to confirm the alignment of both detectors.

\subsection{Off-nominal Calibration}

In addition to the large data sets needed for spectral resolution and quantum efficiency calibration, the flight devices will also require brief data sets in off-nominal conditions.  Charge injection will need to be exercised, to confirm proper function and to calibrate the settings required for a given size of injection charge. Any instrument modes that change the transfer or readout speeds of the detector will be checked, to confirm and quantify any calibration changes. A final optical light leak check will confirm that the on-chip Aluminum optical blocking filter is free of pin holes or other defects.

\section{Conclusion}

AXIS will provide high-throughput, high-spatial resolution X-ray imaging spectroscopy. The AXIS CCDs will go through a ground calibration process which selects the best devices for flight and then accumulates large quantities of data to measure the quantum efficiency and spectral response as a function of energy and position on the CCD.  These are then in turn used to create response products, RMFs and ARFs, which allow scientific analysis.

\acknowledgments % equivalent to \section*{ACKNOWLEDGMENTS}       
 We acknowledge support from NASA for the AXIS Probe Phase A study under contract 80GSFC25CA019, from the Kavli Research Infrastructure Fund of the MIT Kavli Institute for Astrophysics and Space Research, and from the Kavli Institute for Particle Astrophysics and Cosmology at Stanford University. 

% References
\bibliography{report} % bibliography data in report.bib
\bibliographystyle{spiebib} % makes bibtex use spiebib.bst

\end{document}